%% file: main.tex
\input{config/paperMetaData.tex}

\input{config/journalConfigs/laPreprint.tex}
\input{config/journalConfigs/laPreprint_customization.tex}

\newcommand{\qSepSpace}{\vspace{7pt}}

\input{config/commonConfigStuff.tex}

\newcommand{\excludeForCount}[1]{}

\usepackage[sort&compress, numbers, square]{natbib}

\begin{document}
\input{config/journalConfigs/laPreprint_opening.tex}

\begin{abstract}
\input{content/abstract.tex}
\end{abstract}
\section{Introduction}
\input{content/introduction.tex}
\input{content/highlights.tex} 
\input{content/box_perceptualMultistablity}    
\input{captions/figure1.tex}
\input{content/glossary.tex}

\section{Perceptual multistability in psychiatric and neurodevelopmental disorders} \label{sec:perc-mult-psych}
\input{content/multistablityInPsychCondition.tex}

\input{content/table_psych.tex}

\section{Computational psychiatry approaches}\label{sec:comp-psych-appr}
\input{content/compPsychIntro.tex}
\subsection{Bayesian accounts of perceptual multistability} \label{sec:helmh-acco-perc}
\input{content/helmholtzianAccount.tex}

\input{content/box_helmholtzian.tex}

\subsection{Reinforcement learning accounts of perceptual multistability} \label{sec:rl-account}
\input{content/rlAccount.tex}
\input{content/box_pomdp}

\section{Toward a unified synthetis} \label{sec:prospect}
\input{content/prospect.tex}

\section{Concluding remarks}
\input{content/conclusionRemark.tex}
\input{content/outstandingQuestions.tex}
\section{Acknowledgments}
\input{content/acknowledgment.tex}

% \printbibliography
% \input{main.bbl}

\nolinenumbers
\bibliographystyle{unsrtnat} % works for table
\bibliography{locallib4arxiv.bib} 

% DON'T EDIT. If "endfloat" option is enabled all floats appear before appendices
\if@endfloat\clearpage\processdelayedfloats\clearpage\fi

%%%%%%%%%%%%%%%%%%%%%%%%%%%%%%%%%%%%%%%%%%%%%%%%%%%%%%%%%%%%
%%% SUPPLEMENTARY MATERIAL / APPENDICES
%%%%%%%%%%%%%%%%%%%%%%%%%%%%%%%%%%%%%%%%%%%%%%%%%%%%%%%%%%%%
\newpage
% \part*{Appendix}\label{part:appendix}
% \section{Method details}

% label figures here as supp figures
\renewcommand{\figurename}{\bf{Supplementary Figure}}
% reset the counter
\setcounter{figure}{0}  
\setcounter{table}{0}

\pagebreak
% \newpage
% \section{Supplementary figures}
% \input{content/supp_figs.tex}

%% Sadly, we can't use floats in the appendix boxes. So they don't "float", but use \captionof{figure}{...} and \captionof{table}{...} to get them properly caption.

% \begin{appendix}

% \begin{appendixbox}\label{app:ttt}
%     \input{src/supplementary/appendices.tex}
% \end{appendixbox}

% \begin{appendixbox}
%     \input{src/supplementary/resources.tex}
% \end{appendixbox}

% \end{appendix}

%%%%%%%%%%%%%%%%%%%%%%%%%%%%%%%%%%%%%%%%%%%%%%%%%%%%%%%%%%%%
%%% ARTICLE END
%%%%%%%%%%%%%%%%%%%%%%%%%%%%%%%%%%%%%%%%%%%%%%%%%%%%%%%%%%%%

\end{document}

%% file: config/paperMetaData.tex
\newcommand{\paperTitle}{Perceptual multistability: a window for a multi-facet understanding of psychiatric disorders}

\newcommand{\authorone}{Shervin Safavi}
\newcommand{\authortwo}{Danaé Rolland}
\newcommand{\authorfive}{Philipp Sterzer}
\newcommand{\authorthree}{Renaud Jardri}
\newcommand{\authorfour}{Pantelis Leptourgos}

\newcommand{\afillA}{Computational Neuroscience, Department of Child and Adolescent Psychiatry, Faculty of Medicine, TU Dresden, Germany}
\newcommand{\afillB}{Department of Computational Neuroscience, Max Planck Institute for Biological Cybernetics, Germany}
\newcommand{\afillC}{University of Lille, Centre Lille Neuroscience and Cognition, INSERM U1172, CHU Lille, Lille, France}
\newcommand{\afillD}{University of Basel, Department of Psychiatry, Basel 4001, Switzerland}

%% file: config/journalConfigs/laPreprint.tex
\documentclass[9pt]{lapreprint}
\usepackage[version=4]{mhchem} % For chemical notation
\usepackage{siunitx}    % For SI units
\usepackage{pdflscape}  % For putting pages in landscape mode
\usepackage{rotating}   % For rotating specific elements
\usepackage{textgreek}  % Greek symbols
\usepackage{gensymb}    % Symbols
\usepackage[misc]{ifsym} % For the \Letter symbol
\usepackage{orcidlink}  % For the \orcidlink
\usepackage{listings}   % For inserting code chunks
\usepackage{colortbl}   % For Knitr table colouring
\usepackage{tabularx}   % For making Knitr tables compatible
\usepackage{longtable}  % For multi-page tables
\usepackage{subcaption}
\usepackage{multirow}
\usepackage{snotez}     % For sidenote environments. enotez for endnotes
\usepackage{csquotes}   % For language-based quote rules (helps BiBLaTeX)

\DeclareSIUnit\Molar{M}

\title{\paperTitle}

\author[ \orcidlink{0000-0002-2868-530X} 1,2 \Letter]{\authorone} % SS
\author[ \orcidlink{0009-0003-9585-8842} 3]{\authortwo} % DR
\author[ \orcidlink{0000-0003-4687-2317} 4]{\authorfive} % PS
\author[ \orcidlink{0000-0003-4596-1502} 3 \Letter]{\authorthree} % RJ
\author[ \orcidlink{0000-0002-0801-5501} 3]{\authorfour} % PL

\affil[1]{\afillA} % Dresden
\affil[2]{\afillB} % MPI
\affil[3]{\afillC} % France
\affil[4]{\afillD} % Swiss

\correspondence{research@shervinsafavi.org, renaud.jardri@univ-lille.fr}{S. Safavi, R. Jardri}

\leadauthor{Safavi}
\shorttitle{Perceptual multistability: a window for understanding brain dysfunctions}

%% file: config/journalConfigs/laPreprint_customization.tex
\newcommand{\mfig}[1]{\autoref{#1}}

%% file: config/commonConfigStuff.tex
\usepackage{xfrac}

\usepackage{paralist}

\usepackage{xspace}
\usepackage{tabularx}
\usepackage{multirow}
\usepackage{array} 
\usepackage{MnSymbol}
\usepackage{booktabs}
\usepackage[textsize = tiny,backgroundcolor=blue!8!white]{todonotes}
\newcommand{\cut}[1]{}

\newcommand{\shsc}[1]{}
\newcommand{\shsd}[1]{}
\newcommand{\shstd}[1]{}

\newcommand{\olishd}[1]{} % outlineish done

%% file: config/journalConfigs/laPreprint_opening.tex
\maketitle

%%% Local Variables:
%%% mode: latex
%%% TeX-master: "main"
%%% End:

%% file: content/abstract.tex
Perceptual multistability, observed across species and sensory modalities, offers valuable insights into numerous cognitive functions and dysfunctions.
For instance, differences in temporal dynamics and information integration during percept formation often distinguish clinical from non-clinical populations. 
Computational psychiatry can elucidate these variations, through two primary approaches: (i) Bayesian modeling, which treats perception as an unconscious inference, and (ii) an active, information-seeking perspective (e.g., reinforcement learning) framing perceptual switches as internal actions.
Our synthesis aims to leverage multistability to bridge these computational psychiatry subfields, linking human and animal studies as well as connecting behavior to underlying neural mechanisms. 
Perceptual multistability emerges as a promising non-invasive tool for clinical applications, facilitating translational research and enhancing our mechanistic understanding of cognitive processes and their impairments.

\vspace{6pt}
\textcolor{gray}{\textbf{Keywords:}}\\
Bistability;
Computational psychiatry;
Neurodevelopment;
Bayesian inference;
Reinforcement learning;
Neural mechanisms

%% file: content/introduction.tex
Perceptual multistability corresponds to the dynamic alternation of perception that arises when an ambiguous sensory input has several interpretations (\autoref{sec:box:multistability}).
This phenomenon has been studied for decades \citep{blakeVisualCompetition2002a,brascampLawsBinocularRivalry2015,brascampMultistablePerceptionRole2018,blake2022perceptual,safaviMultistabilityPerceptualValue2022},
across a variety of species (\autoref{fig:mspForIntegration}a)
and in various sensory modalities (\autoref{fig:mspForIntegration}b).  
It has been linked to a diverse set of cognitive functions, including 
consciousness \citep{sethTheoriesConsciousness2022},
perception \citep{brascampMultistablePerceptionRole2018,janbrascampNegligibleFrontoparietalBOLD2015},
attention \citep{maierGrowingEvidenceSeparate2021},
decision-making \citep{krugCodingPerceptualDecisions2020}, 
foraging \citep{safaviMultistabilityPerceptualValue2022},  
interoception \citep{vaninteroception,veillette2024cardiac}, and language \citep{styrnal2023bank}.
This makes multistability an ideal paradigm for opening an integrative window of understanding on normal and aberrant cognition.

We believe that the multifaceted nature of perceptual multistability (\autoref{fig:mspForIntegration}) can be used to approach psychiatric and neurodevelopmental disorders, which are regularly accompanied by a wide range of complex cognitive symptoms. Indeed, several aspects of perceptual multistability differ in a wide variety of clinical conditions, however, an integrative synthesis of the findings remains elusive which is discussed in the following.

%%% Local Variables:
%%% mode: latex
%%% TeX-master: "../main"
%%% End:

%% file: content/highlights.tex
\begin{highlightbox}
\caption{Highlights}
\label{sec:highlights}

Perceptual multistability is a decades-old phenomenon that has inspired a wealth of findings in neuroscience and psychology and has great potential to provide an integrative understanding of brain functions and dysfunctions.

\qSepSpace

It has great potential for understanding psychiatric disorders, as differences in perceptual inference and dynamics are observed across a broad range of clinical populations. Furthermore, it can synergize with recent advancements in computational psychiatry, in particular, transdiagnostic approaches.

\qSepSpace

Novel computational frameworks can integrate classical Bayesian and recent reinforcement learning approaches, and provide a framework to explain these differences and bridge cognitive and neural mechanisms. 

\qSepSpace

Given the translational potentials of multistability it can also be a promising experimental tool to link malfunctioning computations to underlying neurobiology in animal models of psychiatric disorders.

\end{highlightbox}

%% file: content/box_perceptualMultistablity.tex
\begin{featurebox}
\caption{Perceptual multistability}
\label{sec:box:multistability}

Perceptual multistability is a phenomenon in which exposure to an ambiguous sensory input leads to more than one percept. Bistability is a specific case of multistability in which spontaneous perceptual alternation (\emph{switches}) focuses on only two distinct interpretations. There are many forms of bistable paradigms, such as 
(i) \emph{ambiguous figures}, which yield shifting interpretations, like the Necker cube or the Schroeder stairs ; (ii) \emph{structure-from-motion stimuli} (SFM), where the perceived direction of motion alternates due to stimulus ambiguity ; or 
(iii) \emph{Binocular rivalry}, where two different images are presented simultaneously and continuously to each eye.
Binocular rival stimuli can also involve \emph{piecemeal perception}, where percepts mix dynamically, sometimes propagating as perceptual traveling waves \citep{leeTravelingWavesActivity2005}. 
This has not been observed with ambiguous figures, as the two interpretations are mutually exclusive in that case. 
To better control perceptual switching experimentally, alternative paradigms have been developed, such as 
various forms of flash suppression \citep{sterzerAccessEmotionalInformation2011,tsuchiyaNoReportParadigmsExtracting2015}, where one stimulus suppresses the other upon sudden presentation of a competing stimulus \citep[flash suppression, ][]{wolfe1984reversing} or presentation of a dynamic noise pattern \citep[continuous flash suppression or CFS,][]{tsuchiya2005continuous}.

Some metrics are commonly used to characterize these perceptual alternations, such as \emph{dominance duration} and switch rate. 
While binocular rivalry is one of the most studied paradigms \citep{blakeVisualCompetition2002}, behavioral evidence suggests that the underlying mechanisms are partly shared between the three paradigms. In particular, they exhibit similar temporal dynamics \citep{brascampDistributionsAlternationRates2005, vaneeDynamicsPerceptualBistability2005}, with dominance durations typically following gamma-like distributions \citep{pastukhovMultistablePerceptionBalances2013a,brascampDistributionsAlternationRates2005} and exhibiting scale-free fluctuations \citep{gaoInertiaMemoryAmbiguous2006a,bakouieScalefreenessDominantPiecemeal2017}. 
In addition, the effects of intermittent blank periods on perceptual dynamics are comparable between the three types of tasks \citep{leopoldStablePerceptionVisually2002}, as well as the psychophysiological relationships between stimulus strength and alternation rates \citep{klinkEarlyInteractionsNeuronal2008a}.
Various factors have been reported to modulate dominance durations, including stimulus contrast \citep[see Levelt's propositions, ][]{brascampLawsBinocularRivalry2015} and perceptual training \citep{suzukiLongTermSpeedingPerceptual2007,klinkExperienceDrivenPlasticityBinocular2010}. Crucially, voluntary control over switching remains limited \citep{scocchiaTopdownInfluencesAmbiguous2014}, but extensive training for a task can bias dominance towards trained features \citep{dieterPerceptualTrainingProfoundly2016}.

Although there are many commonalities among the bistability paradigms, some differences can also be highlighted. For instance, binocular rivalry appears more robust to voluntary control than ambiguous figures \citep{mengCanAttentionSelectively2004}. Importantly, when comparing findings across paradigms, switch rates induced by rival stimuli and ambiguous figures tend to be highly correlated within individuals \citep{shannonGenesContributeSwitching2011, patelIndividualDifferencesTemporal2015}, but a more limited association is found between switch rates observed in binocular rivalry and those in SFM paradigms \citep{caoIndependentSharedMechanisms2018a}, suggesting some differences in the underlying mechanisms.

\end{featurebox}

%% file: captions/figure1.tex
\begin{figure}[hpt]
  \vspace*{-.9cm}
  \centering
  \includegraphics[width=\linewidth]{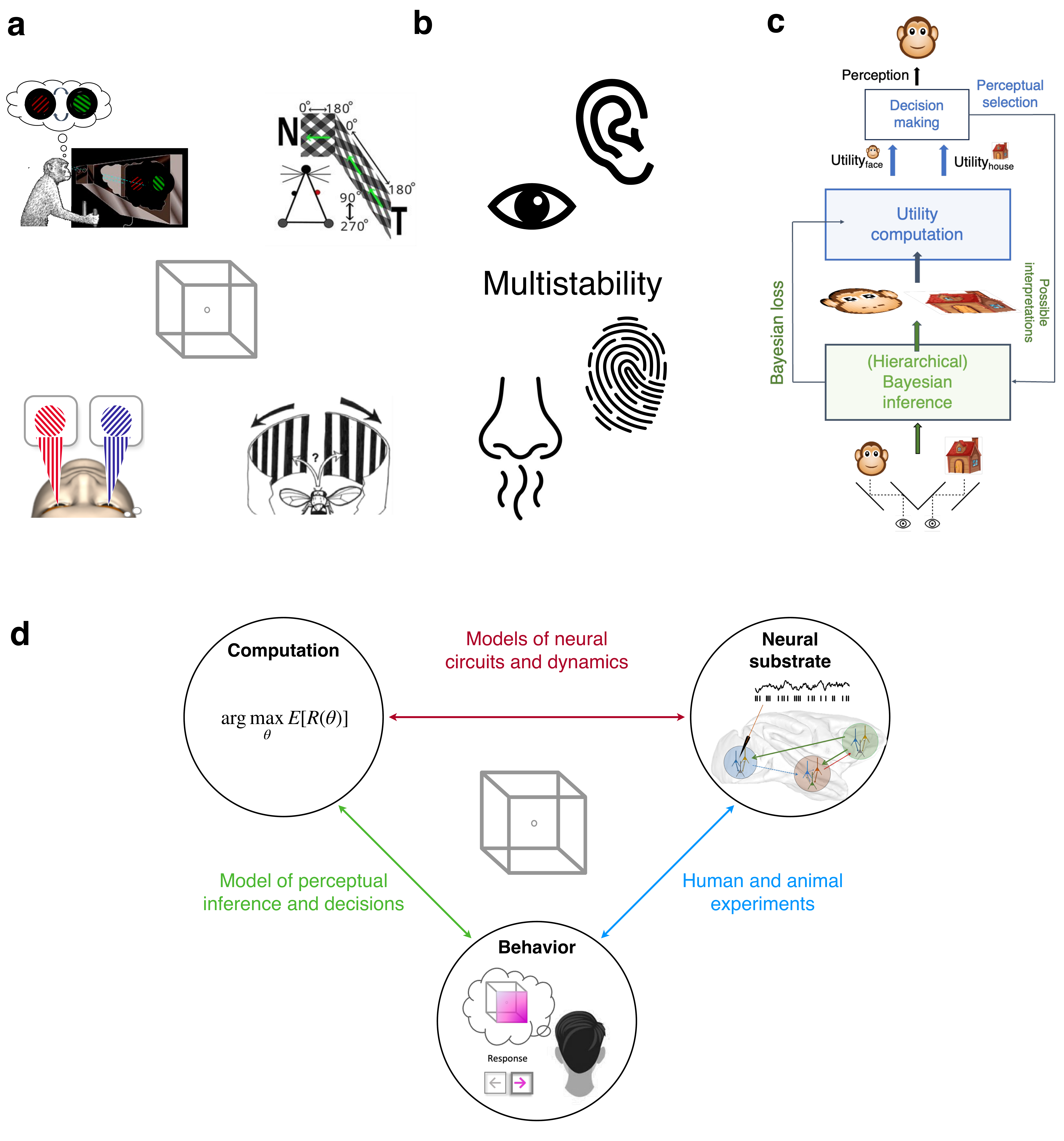}
  \caption{
    \textbf{Perceptual multistability for an integrative understanding of brain functions and dysfunctions.}\\
    Perceptual multistability has been studied
    \textbf{(a)}
    across a wide variety of species, 
    including 
    Drosophila \citep{toepferMultistabilityAmbiguousVisual2018a},
    fish and reptiles \citep[][]{carterPerceptualRivalryAnimal2020b},
    mice \citep{palaginaComplexVisualMotion2017}
    primates \citep{blakeVisualCompetition2002a,panagiotaropoulos2024integrative},
    humans  \citep{zaretskayaZoominginHigherlevelVision2021,bergmann2024cortical}, and even machines \citep[][]{gruber2018perceptual,toosi2024illusions};
    \textbf{(b)}
    and in almost all sensory modalities,
    including
    vision \citep{blakeVisualCompetition2002a},
    audition \citep{kondoSeparabilityCommonalityAuditory2012,einhauserUsingBinocularRivalry2017},
    olfaction \citep{zhouBinaralRivalryNostrils2009a},
    somatosensation \citep{darkiPerceptualRivalryVibrotactile2021},
    and the sense of balance \citep{alaisVestibularActiveSelfmotion2021}. 
    \textbf{(c)}
    As in conventional Bayesian accounts of perception, we can consider an inference module where, through a rich (hierarchical) inference machinery, a possible interpretation will be extracted.
    Then, inferred interpretations go through a \emph{valuation} process that determines the utility of each percept.
    Lastly,  human or animal perception will be determined through a decision process.
    \textbf{(d)}
    Perceptual multistability can be a comprehensive task-performing model with the potential to bridge aspects of (mal)functioning cognitive functions (behavior) and computations to underlying circuit-level biophysics.
    In (a) top-right illustration is adapted from \citep{bogatovaTugofpeaceVisualRivalry2023}, 
    bottom-left from \cite{dieter2011understanding},
    bottom-right from \citep{miller2012attentional}.     
    \label{fig:mspForIntegration}
  }
\end{figure}

%%% Local Variables:
%%% mode: latex
%%% TeX-master: "../main"
%%% End:

%% file: content/glossary.tex
\begin{highlightbox}
\caption{Glossary}
\label{hl:glossary}

\noindent\textbf{Binocular rivalry}\\
A special form of perceptual multistability that involves the two eyes of a participant seeing images that are sufficiently different from each other.
Typical perceptual experience during rivalry is spontaneous perceptual switches between images. 

\noindent\textbf{Ambiguous figures}\\ Figures that are compatible with several (often two) interpretations. For example, 2D projections of 3D figures -- such as the Necker Cube, the Necker lattice, and the Schröder staircase -- in which ambiguity arises from the absence of depth cues.

\noindent\textbf{Switch rate}\\
The rate of perceptual alternations upon exposure to binocular rivalry and ambiguous figures, which is a common measure of (in-)stability, and exhibits substantial variability between individuals but tends to remain stable within individuals over time.

\noindent\textbf{Neurodevelopmental disorders}\\
A group of conditions that originate in the developmental period, often causing learning disabilities or cognitive delays that may persist into adulthood. Examples include \emph{Autism Spectrum Disorder} (ASD) and \emph{Attention-Deficit/Hyperactivity Disorder} (ADHD).

\noindent\textbf{Psychotic symptoms}\\
Experiences characterized by a loss of contact with reality, including hallucinations (perceiving nonexistent things), delusions (unshakable beliefs inconsistent with one's culture and background) and disorganized thinking. Psychotic symptoms are part of \emph{Schizophrenia}'s positive dimension, but can also occur in various other clinical conditions.

\noindent\textbf{Perceptual inference}\\
The process by which the brain interprets sensory information, by combining incoming data with prior knowledge.  

\noindent\textbf{Prediction error}\\
The difference between sensory inputs (what actually happens) and priors (what you expect will happen). They are a central component of predictive coding models, where the brain continuously updates its beliefs by minimizing these mismatches. 

\noindent\textbf{Variational inference}\\
A method of approximate inference where, instead of computing an intractable posterior distribution directly, a simpler distribution is chosen and optimized to minimize the difference (typically via KL divergence) between the two. 

\noindent\textbf{Climbing (ascending) and descending loops}\\
Reverberations of sensory inputs (climbing loops) and priors (descending loops), leading to overcounting of information and inflated confidence. These loops are central features of the circular inference framework. 

\noindent\textbf{Gradient descent}\\
An optimization algorithm used to minimize a function by iteratively moving in the direction of the negative gradient (steepest descent). In variational inference, it is used to optimize the parameters of the approximate (variational) distribution so that it closely matches the true posterior.  
 
\noindent\textbf{Reinforcement learning (RL)}\\
A computational framework to explain (and design) goal-driven decision processes in biological (and artificial) agents.  
In RL, we assume an agent learns to make decisions by interacting with an environment to maximize cumulative return. The agent chooses optimal actions by receiving feedback in the form of rewards and punishments. 

\noindent\textbf{Internal actions}\\
Covert cognitive or/and neural processes that influence behavior \emph{without direct external reflection}. Examples of internal actions would be mental simulations, choosing information to relay to working memory, attentional shifts and emotional regulation mechanisms. Unlike (external) actions, internal actions operate within the brain, guiding future behavior or modifying perception.

\end{highlightbox}

%%% Local Variables:
%%% mode: latex
%%% TeX-master: "../main"
%%% End:

%% file: content/multistablityInPsychCondition.tex
\subsection[Percept formation]{Percept formation} \label{sec:psychInfoInteg}

The formation of a percept is the result of integrating several factors, primarily the sensory input, with our priors and emotional and motivational states. Experimental manipulations of either of these components during multistability tasks have been used to investigate altered perceptual mechanisms in clinical populations.  

Priors that influence percept formation can be either 'hard-wired' (e.g., based on regularities in the sensory environment) or induced experimentally \citep{delangeHowExpectationsShape2018, teufelFormsPredictionNervous2020}.
A well-documented perceptual bias that may be considered hard-wired is the predominance of a “seen from above” interpretation of the Necker cube and other 3D figures \citep{dobbinsAsymmetriesPerception3D2010, mamassianBayesianModellingVisual2002}. This bias is attenuated in patients with schizophrenia \citep[SCZ, ][]{calvertPerceptionVisualAmbiguous1988,keilDynamicalAspectsMotor1998}, and individuals with autism spectrum disorder \citep[ASD, ][]{kornmeierDifferentViewNecker2017}, suggesting reduced reliance on hard-wired priors in these disorders. 
To induce priors experimentally, \citet{schmackEnhancedPredictiveSignalling2017} used a placebo-like manipulation to induce expectations about the rotation direction of an ambiguous sphere. A positive correlation between the induced bias and the severity of \textbf{psychotic symptoms} (\nameref{hl:glossary}) and increased functional connectivity between frontal and visual areas was demonstrated in both SCZ patients and non-clinical populations \citep{schmackDelusionsRoleBeliefs2013,schmackEnhancedPredictiveSignalling2017}. 
Thus, various types of perceptual priors (hard-wired and experimentally-induced) differentially affect percept formation in clinical populations, suggesting separate underlying mechanisms. \citep{teufelFormsPredictionNervous2020}

Implicit emotional processing also affects multistable perception when stimuli are emotionally charged. In non-clinical populations, individuals tend to prefer smiling faces during pleasant emotional states and scowling faces during unpleasant ones \citep{andersonWhatYouFeel2011}. Given that psychiatric and neurodevelopmental disorders often involve abnormalities in emotional processing, these biases are expected to be different in these populations. 
Trait anxiety has been associated with a bias in the initial perception of emotional expressions, specifically an increased tendency toward negative emotional expressions and away from happy faces \citep{grayInfluenceAnxietyInitial2009}. An initial perceptual bias toward fearful faces has also been observed in patients with social anxiety, without an overall increase in dominance of fearful but a decreased dominance of smiling faces \citep{singerEyesAnxietyDissecting2012, andersonSmilesMayGo2013}. In major depressive disorder (MDD), using CFS (\autoref{sec:box:multistability}), shorter suppression of sad faces and longer suppression of happy faces was observed \citep{sterzerAccessEmotionalInformation2011}. On the contrary, in subthreshold depression (StD), a global increase in preference for emotional faces, both sad and happy, over neutral faces has been reported. Importantly, this preference correlates with depression severity, and predicts clinical outcomes over a four-month period \citep {qiuStatedependentAlterationsImplicit2023}. While these findings pertain to anxiety disorders and depression, similar impairments may warrant investigation across a broader range of disorders. These results highlight the utility of multistability as a tool for fine-grained analysis of emotional processing, and offer insights into shared deficits across disorders.

Finally, multistable perception can be used to assess the sensitivity to changes in sensory information. When the ambiguity level of SFM stimuli is gradually changed, SCZ patients show an increased proportion of perceptual states congruent with sensory cues compared to controls. This sensory bias was correlated with the participants' propensity to hallucination \citep{weilnhammerPsychoticExperiencesSchizophrenia2020}. An analogous effect of sensory cues was observed in healthy individuals under the NMDA receptor antagonist ketamine, pointing to a glutamatergic mechanism underlying perceptual alterations in SCZ \citep{weilnhammerNmethyldaspartateReceptorHypofunction2025a}. Thus, multistability can capture changes in sensory processing linked to psychotic symptoms dimension. 

\subsection[Percept stability]{Percept stability} \label{sec:psychStability}

Perhaps the most extensively studied aspect of perceptual multistability in clinical populations is perceptual stability, 
\textbf{Switch rate} \citep[\nameref{hl:glossary}, also see, ][]{pettigrewStickyInterhemisphericSwitch1998,millerGeneticContributionIndividual2010,caoIndependentSharedMechanisms2018a}
that may serve as a trait marker for the predisposition toward psychiatric disorders. 
Most prominently, bipolar disorder (BD) has been repeatedly associated with increased perceptual stability \citep[][]{pettigrewStickyInterhemisphericSwitch1998, millerSlowBinocularRivalry2003, vierckFurtherEvidenceSlow2013, nagamineDifferenceBinocularRivalry2009, lawEvidenceThatEyemovement2017, yeSlowerLessVariable2019} during \textbf{binocular rivalry} (\nameref{hl:glossary}). This has been reproduced using \textbf{ambiguous figures} (\nameref{hl:glossary}), the Necker Cube \citep{huntFluctuationAmbiguousFigure1933}, and SFM stimuli \citep{krugPerceptualSwitchRates2008}.
Increased stability was also shown in patients with MDD using binocular rivalry \citep{jiaDifferenceBinocularRivalry2015, jiaDifferenceBinocularRivalry2020, meldmanQuantitativeAnalysisAnxiety1965, yeSlowerLessVariable2019}, \citep[but see, ][]{millerSlowBinocularRivalry2003}, using the Necker Cube \citep{meldmanQuantitativeAnalysisAnxiety1965}, and SFM \citep{araniBistablePerceptionDiscriminates2024}. 
Finally, increased stability has been reported in SCZ patients and their first-degree relatives using binocular rivalry \citep{foxRateBinocularRivalry1965, wrightBinocularRivalrySlower2003a, xiaoSlowBinocularRivalry2018, yeSlowerLessVariable2019} \citep[but see,][]{millerSlowBinocularRivalry2003}. 
However, in SCZ patients, evidence from other paradigms is mixed. Decreased stability was observed with the Rubin vase \citep{keilDynamicalAspectsMotor1998}, whereas no difference from controls was found with the Schröder staircase \citep{calvertPerceptionVisualAmbiguous1988, keilDynamicalAspectsMotor1998}. Furthermore, SFM experiments have shown reduced perceptual stability for intermittent stimulus presentations in SCZ patients, while results are inconsistent for continuous presentations \citep{schmackPerceptualInstabilitySchizophrenia2015, araniBistablePerceptionDiscriminates2024,albertHierarchicalStochasticModel2017a}. Interestingly, a recent study used a continuously presented SFM stimulus to investigate perceptual stability along the psychosis spectrum -- in SCZ, schizoaffective disorder, BP with psychotic features -- and revealed faster switch rates both in patients and their first-degree relatives \citep{killebrewFasterBistableVisual2024}.
Reduced stability was evidenced in generalized anxiety disorder (GAD), correlating with both trait and state anxiety \citep{jiaDifferenceBinocularRivalry2020,andersonSmilesMayGo2013}, and with anxious personality in the general population \citep{nagamineAcceleratedBinocularRivalry2007}.

Overall, compelling evidence indicates increased perceptual stability in binocular rivalry across various psychiatric conditions (with the exception of anxiety disorders). 
This supports a transdiagnostic perspective and suggests shared genetic underpinnings \citep{millerGeneticContributionIndividual2010,ngoPsychiatricGeneticStudies2011a,lawEffectStimulusStrength2017,chenGenomicAnalysesVisual2018}. However, other paradigms have yielded conflicting results, especially in SCZ. These discrepancies may stem from heterogeneity within diagnostic categories and variations in experimental designs.

\subsection[Percept failures]{Percept failures} \label{sec:psychMixPerc}

Binocular rivalry can induce “mixed” percepts (\autoref{sec:box:multistability}), which can be viewed as "percept failures" in the sense that, the perceptual system fails to build a univocal percept from sensory evidence. These mixed percepts have been particularly investigated in neurodevelopmental disorders. 
An increase in mixed percepts, associated with an overall slower switch rate, has been reported in adults with ASD \citep{freybergReducedPerceptualExclusivity2015a, robertsonSlowerRateBinocular2013, robertsonReducedGABAergicAction2016, spiegelSlowerBinocularRivalry2019}. 
Increased mixed percepts were also observed in adults with attention-deficit/ hyperactivity disorder (ADHD), correlating with inattention severity \citep{jusyteBinocularRivalryTransitions2018}. GABA levels in the visual cortex correlates with the suppression of mixed percepts in non-clinical populations \citep {robertsonReducedGABAergicAction2016}, and pharmacological increase of GABAergic signaling reduces the proportion of mixed percepts \citep{mentchGABAergicInhibitionGates2019}. 
These findings suggest that mixed percepts might result from a lack of perceptual suppression due to a disrupted excitatory-inhibitory (E-I) balance.

Importantly, increased mixed percepts were not observed in children with ASD \citep{karaminisBinocularRivalryChildren2017}, whereas children with ADHD exhibit similar patterns to adults \citep{casanovaOnsetTimeBinocular2013}. Altogether, binocular rivalry studies in both ASD and ADHD reveal age-dependent and age-invariant patterns across children and adults, offering a unique window into the dysfunctional mechanisms underlying neurodevelopmental conditions.
Few studies have looked at mixed percepts in other psychiatric disorders. Further investigations with a focus on mixed percepts are thus warranted, especially  in the light of theoretical models suggesting altered E-I balance as a key pathophysiological factor in psychosis \citep{jardriAreHallucinationsDue2016a}.

In summary, there is substantial evidence of abnormalities in various aspects of multistable perception across psychiatric and neurodevelopmental disorders (see, \autoref{tab:psychSum} for a summary). We discussed evidence for: (i) altered integration of information, including hard-wired and learned priors, emotional states and sensory cues; (ii) altered perceptual stability in psychiatric disorders; and (iii) a higher prevalence of mixed percepts in neurodevelopmental disorders.
Despite the heterogeneity of the existing literature, multistable perception offers a precise lens through which a range of perceptual and cognitive mechanisms, and their alterations in clinical populations, can be investigated in a transdiagnostic fashion, which has been suggested as a promising direction in computational psychiatry \citep{gillanCharacterizingPsychiatricSymptom2016,rouaultPsychiatricSymptomDimensions2018a,gagneImpairedAdaptationLearning2020}.  
As we will discuss in the following sections, computational approaches hold promise in clarifying the sometimes contradictory findings by generating quantitative predictions, and by bridging the gap between alterations at the neural circuits and neurotransmitter levels in psychiatric/neurodevelopmental disorders and corresponding behavioral findings.

%% file: content/table_psych.tex
\newcolumntype{Y}{>{\raggedright\arraybackslash}p{0.3\textwidth}}
\definecolor{customGreen}{HTML}{008000}

\begin{table}[]
\centering
\begin{tabularx}{\textwidth}{
   >{\raggedright\arraybackslash}X
   Y@{\hspace{2pt}}>{\centering\arraybackslash}p{1.2cm}
   >{\raggedright\arraybackslash}X@{\hspace{2pt}}>{\centering\arraybackslash}p{1.2cm}
   >{\raggedright\arraybackslash}X@{\hspace{2pt}}>{\centering\arraybackslash}p{1.2cm}  }

Disorders & Percept formation &  & Percept Stability &  & Percept Failures&      \\
\hline \hline
Schizophrenia (\textbf{SCZ}) & - \textcolor{orange}{\ensuremath{\downarrow} hard-wired priors}  \newline -  \textcolor{customGreen}{\ensuremath{\uparrow} learned priors} (\ensuremath{\filledstar} psychotic symptoms).  \newline -  \textcolor{customGreen}{\ensuremath{\uparrow} bias induced by disambiguating cues} (\ensuremath{\filledstar} hallucinations)  & \citep{calvertPerceptionVisualAmbiguous1988, keilDynamicalAspectsMotor1998} \newline \citep{schmackEnhancedPredictiveSignalling2017, schmackDelusionsRoleBeliefs2013} \newline \newline \citep{weilnhammerNmethyldaspartateReceptorHypofunction2025a} & \textcolor{customGreen}{Binocular rivalry: \ensuremath{\uparrow} stability} , \textcolor{red}{mixed evidence in other paradigms}  & \citep{foxRateBinocularRivalry1965, wrightBinocularRivalrySlower2003a, xiaoSlowBinocularRivalry2018, yeSlowerLessVariable2019} \textit{\citep{millerSlowBinocularRivalry2003}} \newline \citep{calvertPerceptionVisualAmbiguous1988, keilDynamicalAspectsMotor1998, schmackPerceptualInstabilitySchizophrenia2015, araniBistablePerceptionDiscriminates2024, albertHierarchicalStochasticModel2017a}    &  &  \\
\hline
Bipolar Disorder (\textbf{BD}) &     &      & \textcolor{customGreen}{\ensuremath{\uparrow} stability} & \citep{pettigrewStickyInterhemisphericSwitch1998, millerSlowBinocularRivalry2003, vierckFurtherEvidenceSlow2013, nagamineDifferenceBinocularRivalry2009, lawEvidenceThatEyemovement2017, yeSlowerLessVariable2019, huntFluctuationAmbiguousFigure1933, krugPerceptualSwitchRates2008}.     &    &      \\
\hline
Major Depressive Disorder (\textbf{MDD}) and Sub-treshold depression (\textbf{StD}) & \textbf{MDD}: \textcolor{orange}{Perceptual suppression: \ensuremath{\uparrow} for happy faces, \ensuremath{\downarrow} for sad ones}  \newline \textbf{StD}: \textcolor{orange}{\ensuremath{\uparrow} emotional dominance (\ensuremath{\filledstar} symptoms severity)}  & \citep{sterzerAccessEmotionalInformation2011} \newline \newline \citep{qiuStatedependentAlterationsImplicit2023} & \textbf{MDD}: \textcolor{customGreen}{\ensuremath{\uparrow} stability}   &\citep{jiaDifferenceBinocularRivalry2015, jiaDifferenceBinocularRivalry2020, meldmanQuantitativeAnalysisAnxiety1965, yeSlowerLessVariable2019, araniBistablePerceptionDiscriminates2024} \textit{\citep{millerSlowBinocularRivalry2003}} &     &      \\
\hline
Generalized Anxiety Disorder (\textbf{GAD}) and Social Anxiety &  \textcolor{orange}{\textbf{GAD}: Bias in initial percept (away from positive emotions and towards negative ones}  ( \ensuremath{\filledstar} trait anxiety). \newline \textbf{Social anxiety}: \textcolor{orange}{\ensuremath{\downarrow} dominance of positive stimuli, bias of the initial percept towards negative emotions.} & \citep{grayInfluenceAnxietyInitial2009} \newline \newline \newline \newline \citep{singerEyesAnxietyDissecting2012, andersonSmilesMayGo2013}  & \textbf{GAD}: \textcolor{customGreen}{\ensuremath{\downarrow} stability} (\ensuremath{\filledstar} trait and state anxiety)  & \citep{jiaDifferenceBinocularRivalry2020, andersonSmilesMayGo2013, nagamineAcceleratedBinocularRivalry2007}   &     &       \\
\hline
Austim Spectrum Disorder (\textbf{ASD}) & \textcolor{orange}{\ensuremath{\downarrow} hard - wired priors}  & \citep{kornmeierDifferentViewNecker2017}       &       &       & \textcolor{customGreen}{\ensuremath{\uparrow} mixed percept} (\ensuremath{\filledstar} symptoms severity)  & \citep{freybergReducedPerceptualExclusivity2015a, robertsonSlowerRateBinocular2013, robertsonReducedGABAergicAction2016, spiegelSlowerBinocularRivalry2019}    \\
\hline
Attention Deficit/Hyperactivity disorder (\textbf{ADHD}) &       &  &       &       & \textcolor{orange}{\ensuremath{\uparrow} mixed percepts} (\ensuremath{\filledstar} inattention severity) & \citep{jusyteBinocularRivalryTransitions2018}  \\
\hline
\end{tabularx}
\vspace{10pt}
\caption{
\textbf{Summary of the main findings discussed along the 3 main categories of \nameref{sec:psychInfoInteg}, \nameref{sec:psychStability}, and \nameref{sec:psychMixPerc}.} 
Key findings relevant for the scope of this study are included in this table, thus, this table is not meant to be exhaustive. \ensuremath{\filledstar} : Correlation with symptoms.  Color code: 
\textcolor{customGreen}{Green: high level of evidence, supported by numerous and/or methodologically strong papers.} 
\textcolor{orange}{Orange: middle level of evidence, shown by a limited number (1-2) of studies.} 
\textcolor{red}{Red: low level of evidence: contradictory results across studies.}. Contradictory references are specified in italic \textit{[ref]}
}
\label{tab:psychSum}
\end{table}

% Color code: 
% Green: high level of evidence, supported by numerous and/or methodologically strong papers
% Orange : middle level of evidence, showed by a few paper and/or contradicted more than once. 
% Red: low level of evidence: contradictory results across studies.

%% file: content/compPsychIntro.tex
% Many of the alterations of multistable perception observed in clinical populations can be explained based on two main frameworks in computational psychiatry: (i) Bayesian accounts that conceptualize perception as inference \citep{deneve2016circular,leptourgos2017can,brascampMultistablePerceptionRole2018,gabhart2023predictive} and (ii) reinforcement learning (RL) accounts that emphasize computationally active, information seeking aspects of perception \citep{martinUsefulMisrepresentationPerception2021,safaviMultistabilityPerceptualValue2022,weilnhammer2023sensory}.
% In the following, we extensively discuss these two frameworks and also propose directions to integrate them.

% We believe that the emerging field of computational psychiatry may offer explanations for the differences observed in the mental health continuum,
% % and more importantly, may integrate them into a unified framework.

%% file: content/helmholtzianAccount.tex
Prevalent computational and algorithmic-level accounts of multistable perception trace back to Helmholtz’s seminal suggestion \citep{vonhelmholtzHandbuchPhysiologischenOptik1867} that perception involves a form of unconscious inference (\textbf{perceptual inference}, \nameref{hl:glossary}): our experiences are not pure reflections of sensory inputs, but dynamic interpretations shaped by prior knowledge. In modern terms, perception can be viewed as statistical inference in a generative model of the sensory world \citep{doyaBayesianBrainProbabilistic2007}, where likelihoods and priors are combined according to Bayes’ theorem to yield posterior beliefs about the causes of sensory inputs (\autoref{sec:box-1:-predictive}).

Multistability, with its characteristic phenomenology and apparent dissociation between stimulus and percept, naturally lends itself to Bayesian interpretations. These accounts offer a middle ground between purely feedforward and purely feedback models of perception, by emphasizing the interplay between sensory evidence and priors \citep{brascampMultistablePerceptionRole2018}.

While posterior estimation is straightforward in toy models, it becomes computationally intractable for real-world applications. To address this, computer scientists and neuroscientists have developed various shortcuts which approximate the exact posterior probability distribution \citep{bishop2006pattern}. The same challenge also applies to Bayesian models of multistable perception. We can categorize Bayesian models of multistability into three classes, based on their algorithmic implementation of probabilistic inference.

One influential line of work proposes that the brain approximates the posterior over hidden causes by drawing a set of (correlated) samples \citep{moreno-boteBayesianSamplingVisual2011c,sundareswara2008perceptual}, implementing an approximate Bayesian inference algorithm known as Gibbs sampling \citep{reichert2011neuronal,samuelgershmanPerceptualMultistabilityMarkov2014a}. In this framework, dominant percepts are interpreted as random samples drawn from the underlying posterior distribution. Crucially, this perspective does not assume that the brain maintains a full representation of the posterior; rather, it samples percepts over time, accounting for the stochasticity and temporal structure of perceptual alternations.

A second class of models adopts an \emph{analysis-by-synthesis} approach \citep{dayanHierarchicalModelBinocular1998}, leveraging \textbf{variational inference} (\nameref{hl:glossary}) to approximate the true posterior (\autoref{sec:box-1:-predictive}). Instead of sampling directly from a complicated posterior distribution, it is assumed that the brain optimizes a simpler distribution (often parameterized by its sufficient statistics, such as the mean and variance) to match the true posterior. Within hierarchical predictive coding accounts of cortical function \citep{friston2009predictive}, the brain operates as a statistician, generating and testing hypotheses about the causes of the sensory inputs \citep{lochmann2011neural}. More particularly, predictions generated at higher levels of the hierarchy are compared against lower-level sensory inputs, and the mismatch between the two gives rise to (precision-weighted) \textbf{prediction error} (\nameref{hl:glossary}) signals which are propagated upwards to update beliefs \citep{friston2005theory}. Multistable perception arises when multiple hypotheses about the cause of the input have both high likelihood and high prior probability \citep{hohwy2008predictive,weilnhammer2017predictive}. Since only one object can occupy any spatiotemporal position, the brain must choose among these competing but plausible hypotheses. However, each dominant hypothesis will generate predictions that explain away only part of the sensory input, leaving residual prediction errors. These errors accumulate, destabilizing the current percept and eventually prompting a perceptual switch. A model-based fMRI in conjunction with transcranial magnetic stimulation suggests that prediction errors are represented in bilateral inferior frontal gyri and insulae, favoring a causal role of frontal areas in resolving multistability \citep{zaretskayaIntrospectionAttentionAwareness2014,safaviFrontalLobeInvolved2014,odegaardShouldFewNull2017,safaviNonmonotonicSpatialStructure2018,blockWhatWrongNoReport2019,weilnhammer2021active,michelConsciousPerceptionPrefrontal2022,kapoorDecodingContentsConsciousness2020,dwarakanathBistabilityPrefrontalStates2023}.

While sampling and predictive coding theories assume that the brain approximates the posterior distribution either by generating a set of samples or by computing the sufficient statistics of a simpler, tractable distribution, 
a third view -- the circular inference framework -- suggests that probabilistic computations rely on an imperfect execution of exact inference \autoref{sec:box-2:-circular}. Like predictive coding, circular inference assumes that the brain constructs hierarchical generative models of the environment \citep{jardri2013circular,deneve2016circular}. However, rather than using top-down predictions to explain away bottom-up sensory inputs, top-down and bottom-up messages are integrated. To prevent redundant information from corrupting inference, the system requires control mechanisms that subtract repeated messages. These control mechanisms might be mediated by inhibitory connections, with their activity carefully balanced against excitation.
When these mechanisms are compromised \citep[e.g., due to an E/I imbalance, ][]{jardri2016hallucinations}, recurrent information \textbf{loops} (ascending and descending, \nameref{hl:glossary}) emerge (ascending-amplifying inputs, or descending-amplifying priors), leading to the overcounting of information and extreme posterior estimates (i.e., overconfidence). While the circular inference framework was originally developed as a theory of psychotic symptoms \citep{jardri2013circular,jardri2017experimental}, more recent evidence suggests that mild form of circular inference may be present in at-risk individuals for psychosis \citep{derome2023functional}, but also in the general population \citep{jardri2017experimental,leclercq2024conspiracy}, causally contributing to multistable perception \citep{leptourgos2020circular}. Theoretically, when there are 2 probable but incompatible interpretations of the data, descending loops – reflecting overcounted priors – change the dynamics of belief updating, giving rise to a bistable attractor \citep[i.e., bistable perception,][]{leptourgosFunctionalTheoryBistable2020b}.

It is important to highlight that Bayesian accounts of multistable perception reconceptualize percept choice and alternation as an epistemic response to the apparent incompatibility of the overlapping explanations \citep{hohwy2008predictive}. As a result, they can shed new light on the traditional descriptions of multistable perception in terms of inhibition, adaptation and noise \citep{decoBrainMechanismsPerceptual2013,braunAttractorsNoiseTwin2010,caoBinocularRivalryReveals2021a}. For example, within a sampling framework, neuronal adaptation can be interpreted as a mechanism that facilitates exploration of the posterior, preventing the system from getting stuck in a single mode and thus improving sampling-based inference \citep{reichert2011neuronal}. In the same vein, hierarchical inference reconceptualizes lateral inhibition as competition between top-down explanations, rather than between inputs \citep{dayanHierarchicalModelBinocular1998,hohwy2008predictive,leptourgosFunctionalTheoryBistable2020b}. Finally, noise is no longer seen as an epiphenomenon, but as a fundamental feature of the perceptual process \citep{samuelgershmanPerceptualMultistabilityMarkov2014a,leptourgosFunctionalTheoryBistable2020b}. 

Consequently, Bayesian models can provide functional explanations for the behavioral differences observed between clinical and non-clinical populations. For instance, attenuated prior expectations, as proposed by the strong input (weak priors) theory of psychosis \citep{kapur2003psychosis,sterzer2018predictive}, as well as the circular inference model with ascending loops \citep{jardri2013circular}, may account for the diminished influence of hard-wired priors in SCZ and ASD, along with the heightened sensitivity to external cues \citep[though see, ][]{leptourgosFunctionalTheoryBistable2020b} and possibly enhanced learning. However, other phenomena, such as increased perceptual stability in binocular rivalry tasks, are less readily explained by strong inputs. Instead, these findings align more closely with the overweighting of priors, as posited by the strong prior theory of psychosis \citep{corlett2019hallucinations}. This discrepancy underscores the challenge posed by the heterogeneity inherent in current psychiatric diagnostic categories and highlights the value of computational phenotyping in addressing these complexities. 

Lastly, many aspects of brain functions are better understood as mechanisms that enable an agent to modify its environment \citep[][, also see, \nameref{sec:rl-account}]{dayan2008decision}. Inference models can also incorporate active inference mechanisms, wherein actions can also minimize expected prediction errors \citep[][\autoref{sec:box-1:-predictive}]{friston2005theory}. This self-evidencing implies that the brain not only passively interprets inputs but also acts to sample evidence that confirms its own model. In the context of visual multistability, this includes eye movements that redirect attention toward features favoring a new interpretation, thereby enabling a perceptual switch \citep{parr2019perceptual,novickyBistablePerceptionPrecision2022a}. These actions are believed to be orchestrated by precision control mechanisms, potentially mediated by neuromodulators such as acetylcholine, dopamine, and noradrenaline \citep{novickyBistablePerceptionPrecision2022a}, also involved in perceptual multistability \citep{sheyninCholinergicModulationBinocular2020} \citep{pfefferCatecholaminesAlterIntrinsic2018}
and in many psychiatric disorders \citep{iglesiasModelsNeuromodulationComputational2017a}.

%% file: content/box_helmholtzian.tex
\begin{featurebox}
\caption{Predictive coding}
\label{sec:box-1:-predictive}

The brain is increasingly understood as a Bayesian inference machine, continuously integrating sensory inputs with prior beliefs to infer the hidden causes of the inputs. Formally, this inference process is described by Bayes' theorem:

\begin{equation}
P(X|S)=\frac{P(S|X )\times P(X)}{P(S)}
\end{equation}

where \(X\) denotes the hidden cause of an observed variable \(S\), \(P(X|S)\) the posterior probability, \(P(S|X)\) the likelihood function, \(P(X)\) the prior distribution, and \(P(S)=\int P(S|X)P(X)dX \)  the marginal likelihood or evidence.

In the simple case where both the prior and the likelihood are Gaussian distributions, and under a maximum a posteriori decision rule, Bayesian filtering simplifies to:

\begin{equation}
\mu_{post}=\mu_{pr}+\frac{\sigma_{pr}{^2}}{\sigma_{pr}{^2}+\sigma_{s}{^2} }(\mu_s-\mu_{pr})
\end{equation}

Here, the posterior mean, that is, the brain's updated belief about the hidden cause, is given by the prior mean plus a precision-weighted prediction error, the mismatch between the observed input \(\mu_s\) and the prediction \(\mu_p{_r}\), scaled by the precision (inverse variance) of the prior and sensory input. This simple, yet powerful, update rule forms the basis of predictive coding, originally introduced in the context of early visual processing to account for non-classical surround effects \citep{raoPredictiveCodingVisual1999}.

Building on this foundation, hierarchical predictive coding models have been proposed as a unifying theory of cortical function \citep{friston2005theory}. These models suggest that each level of a cortical hierarchy generates predictions about the activity in the level below, serving as empirical priors, and receives precision-weighted prediction errors from that lower level. Because exact inference is often computationally intractable (because of the difficulty of marginal likelihood estimation), these models are frequently framed as variational inference. In this view, the brain minimizes variational free energy (that can be decomposed into terms like risk, ambiguity, and novelty), a tractable upper bound on the negative log-evidence (surprisal). Interestingly, under Gaussian assumptions, this free energy reduces to the sum of squared, precision-weighted prediction errors \citep{friston2010free}. Within this framework, neural dynamics are thought to implement \textbf{gradient descent} (\nameref{hl:glossary}) on the free energy, a process that governs not only perceptual inference but also synaptic plasticity and model optimization over time.

More recent Bayesian formulations go a step further by incorporating action into the same computational scheme, giving rise to the theory of active inference \citep{friston2017active,friston2016active}. According to this account, organisms can minimize free energy (and thus prediction error) through two complementary strategies: (i) by updating internal expectations to better match incoming data (perception), and (ii) by acting on the environment to sample sensory data that conforms to prior expectations (action). This dual strategy allows for flexible, adaptive behavior grounded in a single, unified principle (also see, \autoref{sec:box-3:-markov}, for an alternative perspective).

\begin{center}
    \includegraphics[width=.3\linewidth]{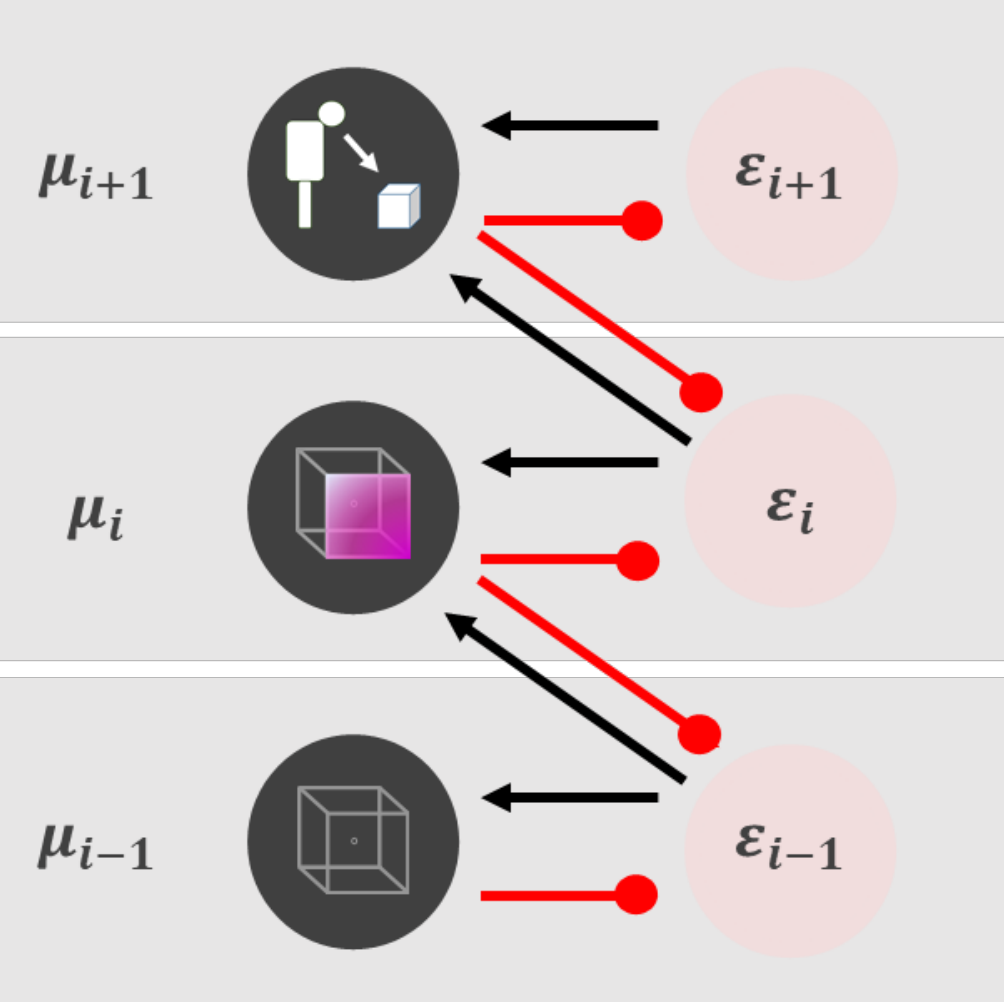}
\end{center}
\vspace{-3pt}
\caption*{
\small
Figure of \autoref{sec:box-1:-predictive}: 
{\normalfont Schematic illustration of predictive processing.}
}

\end{featurebox}

\begin{featurebox}
\caption{Circular inference}
\label{sec:box-2:-circular}

Hierarchical inference can be implemented through a variety of algorithmic frameworks. One of these particularly general and powerful approaches is belief propagation, and its close relative, the sum-product algorithm. These algorithms perform Bayesian inference through local message-passing on directed acyclic graphs, often represented as factor graphs. Unlike predictive coding, in which inference is driven by prediction errors, belief propagation integrates precision-weighted top-down predictions and bottom-up sensory input directly. Crucially, the success of belief propagation depends on a correction mechanism that actively removes redundant information at each level of the hierarchy, specifically that which has previously been communicated in the opposite direction. Without this mechanism, the same piece of evidence can be inadvertently counted multiple times, compromising the accuracy of the inference.

In the special case of binary random variables and pairwise factor graphs (i.e., one cause per variable at most) belief propagation can be recast in a neural-like computational framework. This is exemplified by the model developed by \citet{jardri2013circular}, where message updates and belief estimates follow coupled recursive equations reminiscent of activity in recurrent neural networks:
\begin{equation}
M_{ji}^{(n+1)}=g(B_{i}{^n}-aM_{ij}{^n} )\,,    
\end{equation}
\begin{equation}
B{_i}^{(n+1)}=\sum_jM_{ji}^{(n+1)}\,     
\end{equation}
Here, \(M_{ji}^{(n+1)}\) represents the probabilistic message sent from node j to node i at iteration n+1, and \(B_i^{(n+1)}\) is the posterior belief at node i. The term \(aM_{ij}{^n}\) serves as a correction factor, with the scalar a controlling the extent to which redundant information is removed. When \(a=1\), all redundant information is eliminated, and the algorithm converges to the correct posterior distributions. However, when \(a<1\), residual redundancy can persist, leading to information loops and a failure mode known as \emph{circular inference}, in which messages are overcounted and inference becomes suboptimal.

Circular inference has been proposed as a computational account of aberrant perceptual experiences, including hallucinations, illusions, and synesthesia \citep{notredame2014visual,jardri2013circular,leptourgosFunctionalTheoryBistable2020b}, as well as the formation and maintenance of idiosyncratic beliefs resistant to contradictory evidence \citep{jardri2016hallucinations,leclercq2024conspiracy}. In support of this theory, \citep{jardri2017experimental} used a probabilistic reasoning task to demonstrate that individuals with schizophrenia overcount bottom-up sensory inputs more than control participants, while both groups overcount prior beliefs to a similar extent.

\begin{center}
    \includegraphics[width=.3\linewidth]{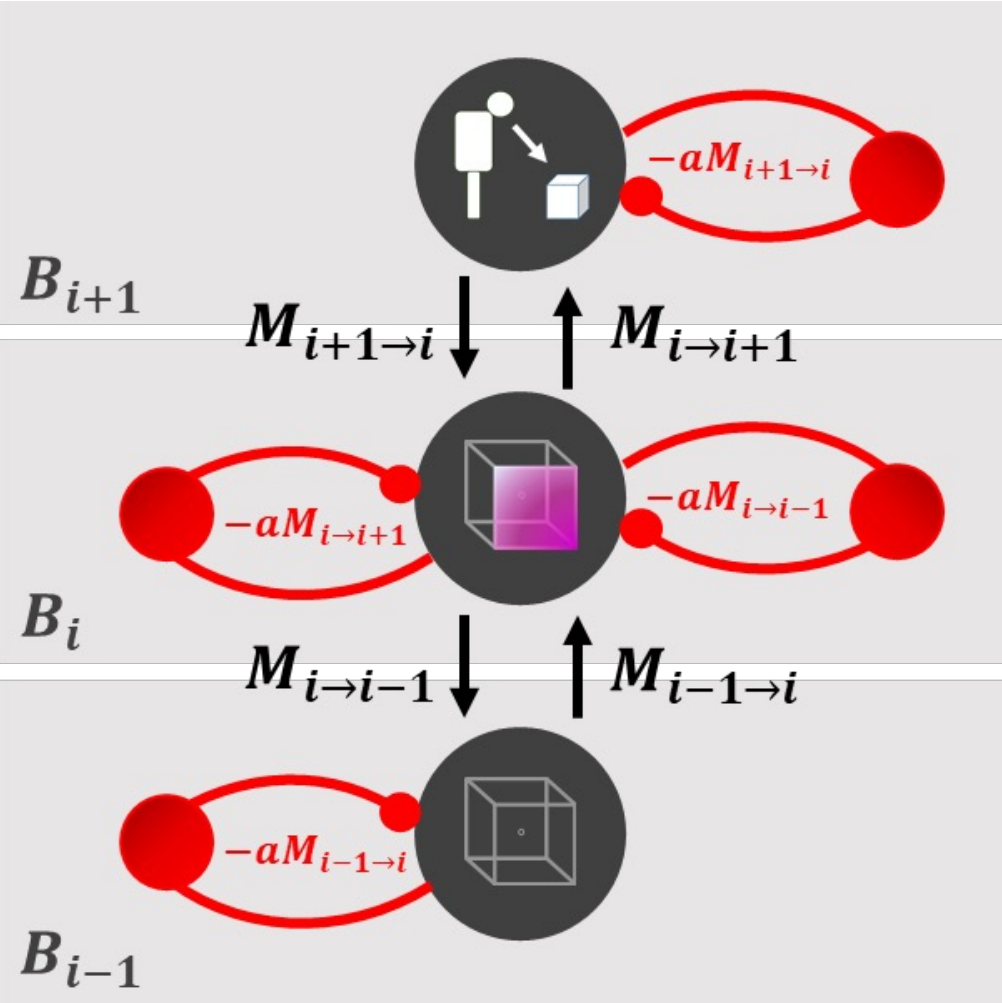}
\end{center}
\vspace{-3pt}
\caption*{
\small
Figure of \autoref{sec:box-2:-circular}: 
{\normalfont Schematic illustration of circular inference.}
}

\end{featurebox}

%% file: content/rlAccount.tex
Recently, complementary to the classical  account of perceptual multistability, computationally active approaches \citep{safaviMultistabilityPerceptualValue2022,safaviDecisiontheoreticModelMultistability2024,martinUsefulMisrepresentationPerception2021a,novickyBistablePerceptionPrecision2022a}, for instance, based on \textbf{reinforcement learning} (\nameref{hl:glossary})
have also been developed.
According to the RL account, spontaneous switches are treated as \textbf{internal actions} (\nameref{hl:glossary}), bringing them under the same umbrella as other decision-making processes \citep{dayanHowSetSwitches2012}.
Such decision-theoretic approaches are not only compatible with the Bayesian account discussed earlier, 
but also offer an explanation for some elusive aspects of multistable experiences, for instance, the modulation of perception by cognitive factors such as reward \citep{balcetisSubjectiveValueDetermines2012,wilbertzReinforcementPerceptualInference2014,marxRewardModulatesPerception2015,wilbertzFMRIbasedDecodingReward2017,pothastUncertainRewardCues2024,lunghiLearnedValueModulates2023,lunghiEffectMonetaryReward2019a,haasCanHierarchicalPredictive2021b,haasCanHierarchicalPredictive2021b},  
emotion \citep{bannermanBinocularRivalryWindow2011}, etc. \citep{safaviMultistabilityPerceptualValue2022} and the apparent (and controversial) involvement of regions such as the prefrontal cortex \citep{zaretskayaIntrospectionAttentionAwareness2014,safaviFrontalLobeInvolved2014,odegaardShouldFewNull2017,safaviNonmonotonicSpatialStructure2018,blockWhatWrongNoReport2019,weilnhammer2021active,michelConsciousPerceptionPrefrontal2022,kapoorDecodingContentsConsciousness2020,dwarakanathBistabilityPrefrontalStates2023} and anterior cingulate cortex \citep{lyuIntrinsicBrainDynamics2022,drewConflictMonitoringAttentional2021,gelbard-sagivHumanSingleNeuron2018} in perceptual switching 
\citep{zaretskayaIntrospectionAttentionAwareness2014,safaviFrontalLobeInvolved2014,tsuchiyaNoReportParadigmsExtracting2015,odegaardShouldFewNull2017,blockWhatWrongNoReport2019,michelConsciousPerceptionPrefrontal2022}.

According to this approach, decisions are made to maximize the expected return. 
More specifically, it is based on the formalism of a \emph{Partially Observable Markov Decision Process} (POMDP) (\autoref{sec:box-3:-markov} and \autoref{fig:mspForIntegration}c). 
While Bayesian approaches conclude with the estimation of the posterior distribution over these interpretations and switching stems from, for instance, sampling from the posterior distribution, in the RL account, this is extended by incorporating a decision-theoretic perspective \emph{that explicitly models the selection between these interpretations as an internal decision} \citep{dayanHowSetSwitches2012,safaviMultistabilityPerceptualValue2022,lubianikerNeurofeedbackLensReinforcement2022a,weilnhammer2023sensory}.
The agent chooses actions (perceptual selections) to maximize cumulative long-run summed utility. 
Utility can be inherent to the stimuli, for example, each image possesses an aesthetic value; or it is \emph{associated} with the percept due to context, for instance, relevance for the task at hand. 
In either case, the utility of the percept can evolve over time, e.g., due to aesthetic boredom or cognitive fatigue \citep[][which may also explain some psychiatric aspects, \nameref{sec:psychMixPerc}]{safaviDecisiontheoreticModelMultistability2024}, or better learning the volatility of the environment \citep{dieterPerceptualTrainingProfoundly2016}.

According to the RL approaches \citep[and some active inference models,][]{hohwy2016distrusting,parr2019perceptual,novickyBistablePerceptionPrecision2022a}, an explicit model of the temporal structure of the environment must be taken into account. 
Crucially, this enables us to explain the temporal dynamics of the perception 
observed in various studies \citep{vaneeDynamicsPerceptualBistability2005,mamassianTemporalDynamicsBistable2005,suzukiLongTermSpeedingPerceptual2007,brascampMultiTimescalePerceptualHistory2008,klinkExperienceDrivenPlasticityBinocular2010a}. 
This is particularly crucial from a psychiatric perspective, since a large category of behavioral phenotypes observed as the difference between the clinical/non-clinical populations concerns the temporal dynamics of perception (\nameref{sec:psychStability}).
Beyond simple formulations based on transition functions, it might be plausible to consider dopamine’s potential role in modulating certain temporal aspects of perceptual switching \citep{vaneeDynamicsPerceptualBistability2005,mamassianTemporalDynamicsBistable2005,suzukiLongTermSpeedingPerceptual2007,brascampMultiTimescalePerceptualHistory2008,klinkExperienceDrivenPlasticityBinocular2010a}.
This hypothesis arises from the idea that tonic dopamine signals encode the local average reward rate, effectively serving as an opportunity cost for time and influencing the speed or urgency of behavioral responses \citep{nivTonicDopamineOpportunity2007,mazzoniWhyDonWe2007,shadmehrMovementVigorReflection2019}.
Intuitively, a high reward rate increases the cost of time spent without reward, encouraging faster decision-making, even at the expense of higher switch rates. Conversely, when rewards are infrequent, the incentive for rapid decisions weakens \citep{nivTonicDopamineOpportunity2007}.
Thus, such RL-based developments can be quite insightful on the psychiatric side.

The RL-inspired study of the role of the dopaminergic system in multistability can provide new insight for understanding both the behavioral pathologies of multistability and its underlying neural circuit dysfunctions.  
Beyond the implications of the dopaminergic system in shaping the temporal dynamics, 
neural dynamics and the coordination of decision-making networks heavily rely on dopaminergic modulation \citep{kahnt2025curious} that can also be quite relevant from a decision-theoretic perspective to multistability.
For instance, dopamine-induced inhibition in the canonical circuit \citep{froudist-walshDopamineGradientControls2021} might account for top-down modulations.
The action of dopamine prevents distractor-related activity from sensory areas disrupting ongoing persistent activity in the frontoparietal network \citep{froudist-walshDopamineGradientControls2021}.
This could be critical in perceptual multistability, as it was shown that the frontoparietal network is critically involved in switching dynamics \citep{brascampMultistablePerceptionRole2018},
and there is evidence supporting the effect of dopaminergic modulation also in perceptual multistability \citep{schmackInfluenceDopaminerelatedGenes2013a,pfefferCatecholaminesAlterIntrinsic2018}.
Thus, the pathological changes in the switch rate can be attributed to dopaminergic modulation influencing the frontoparietal network.
Lastly, the potential role of dopamine also resonates with the circuit mechanisms based on the E-I (im)balance discussed earlier \citep{froemke2015plasticity}.

Lastly, multistability can resonate with the use of cognitive maps, which is a recent RL-inspired development in computational psychiatry \citep{nour2021impaired,nourTrajectoriesSemanticSpaces2023,nour2024cognitive}. 
Multistability may provide a bridge for understanding internal computations that are central to using cognitive maps for planning \citep[][]{nour2021impaired,nourTrajectoriesSemanticSpaces2023,nour2024cognitive}. 
For instance, mental simulations are sophisticated cognitive sequences, however, they are internally generated and also processed, which makes them difficult to access externally \citep{liuDecodingCognitionSpontaneous2022}.  
On the contrary, perceptual switches during multistability could be treated as extremely simplified internally generated sequences with sensible outcomes.
Thus, perceptual multistability could also serve as a crucial tool to understand (malfunctioning) internal computations.

%%% Local Variables:
%%% mode: latex
%%% TeX-master: "../main"
%%% End:

%% file: content/box_pomdp.tex
\begin{featurebox}
\caption{Partially observable Markov decision processes}
\label{sec:box-3:-markov}

Humans and animals (or generically agents) often face situations where they must repeatedly choose one action from multiple options. Each decision in a sequence must consider not only the immediate outcome, such as gaining a reward or altering the agent's state, but also the long-term consequences of these actions. \emph{Markov decision processes} (MDP) provide a general framework for modeling such decision-making scenarios.

An MDP comprises four key components (state, action, transition probability, and reward). To illustrate these, consider the example of an animal navigating its natural habitat.
The relationship between the animal and its environment characterizes the state. For instance, factors that characterize the states are: distances between animals, nutrient sources, and predatory animals. 
More formally, \textit{states} represent the agent's external circumstances.
Then, depending on the situation, the agent can move toward nutrient sources or escape from predators, in both cases, to maximize its survival chance. 
Thus, external and internal \textit{actions} are the choices available to the agent \citep[for latter, see, ][]{safaviMultistabilityPerceptualValue2022,safaviDecisiontheoreticModelMultistability2024}. 
However, taking a given action does not always bring the animals to their desired state. 
For instance, in a windy environment, animal navigation might deviate from its executed action plan. 
More formally, \textit{transition probabilities} characterize how the state changes given each possible action.
Lastly, depending on the state, action, and transition probabilities, the agent will gain (or lose) a \textit{reward}.

While MDPs have been highly effective in modeling various environments, they rely on the restrictive assumption that an agent has complete knowledge of the state at all times, typically through sensory input. This assumption does not hold for many real-world problems, including perceptual multistability.
To address this limitation, a more complex, yet realistic, class of models known as partially observable MDPs (POMDPs) has been introduced. In POMDPs, the agent may be uncertain about the state of the world and must form and maintain a belief about it. The agent's goal is to select actions (or adopt a policy) that maximize cumulative long-term utility based on the available information.
POMDPs and their extensions have been widely used to explain diverse aspects of human and animal behaviour, ranging from decision-making of individuals \citep{hogeveen2022neurocomputational} to social interactions \citep{rusch2020theory}.

\begin{center}
    \includegraphics[width=.85\linewidth]{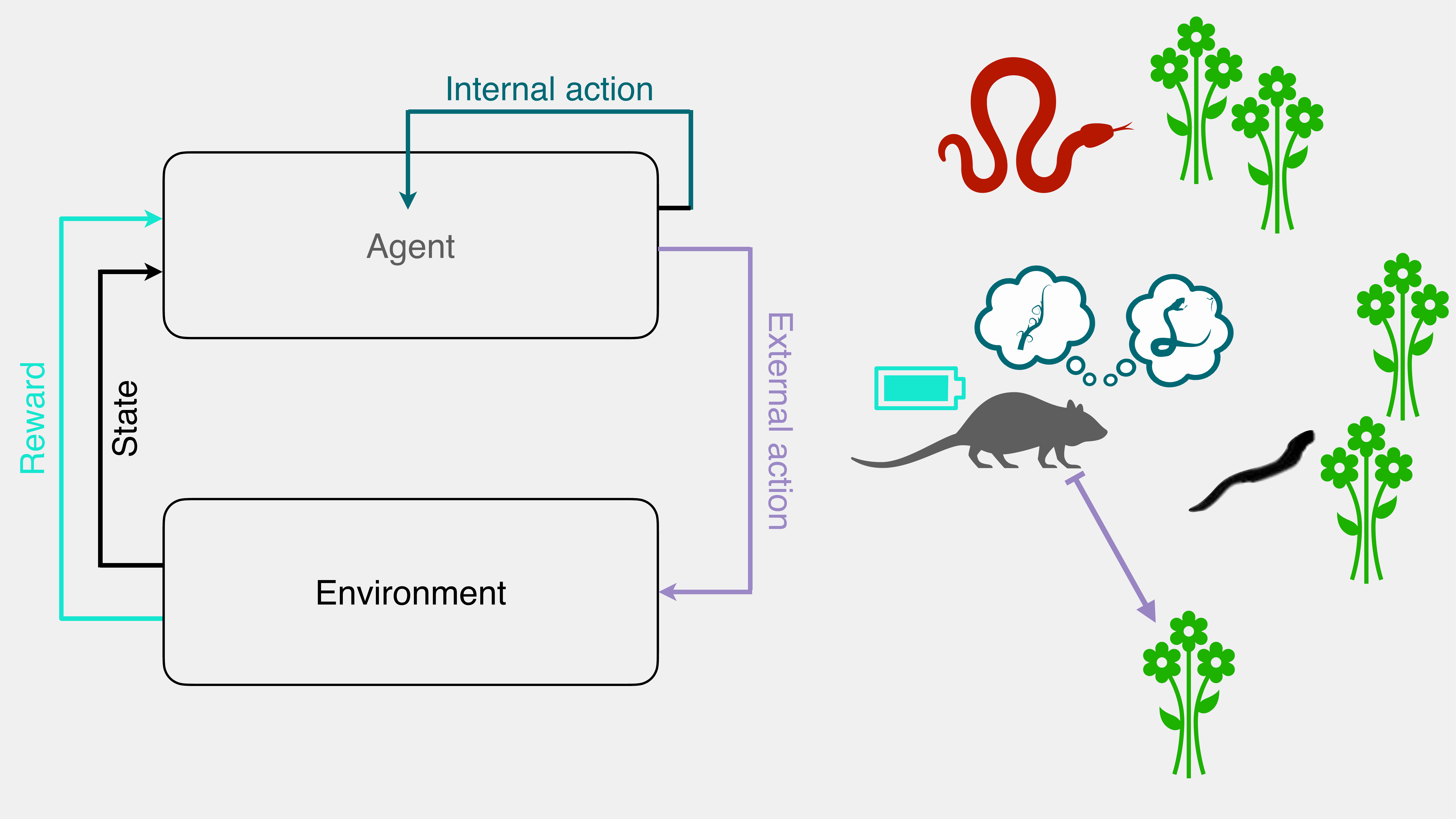}
\end{center}
\vspace{-13pt}
\caption*{
\small
Figure of \autoref{sec:box-3:-markov}: Schematic illustration of POMDP. {\normalfont Left depicts the RL diagram, and right illustrates the agent and its environment. Colors of components on the RL diagram matches the right illustration.}
}

\end{featurebox}

%% file: content/prospect.tex
\subsection{A shared spark across computational descriptions}

In principle, the computations underlying the Bayesian and RL accounts outlined above are intimately related. 
Furthermore, they have been studied extensively to understand how aberrant computations lead to a wide range of cognitive dysfunctions, such as alterations in learning and decision-making \citep{maia2011reinforcement,huysDepressionDecisionTheoreticAnalysis2015,smith2021recent}.
Future approaches should therefore attempt to integrate these two perspectives through extended computational frameworks, as well as experiments designed to evaluate these frameworks.

We propose that there are at least two approaches to integrate Bayesian and RL perspectives on pathology-related observations of multistability. One approach is to expand the RL framework of multistability with richer accounts of belief formation, representation, and maintenance (\autoref{fig:mspForIntegration}c). 
For example, the simplistic Bayesian belief update of POMDP can be replaced by a hierarchical belief update \citep{dayanHierarchicalModelBinocular1998}.
This makes RL models computationally richer and allows us to incorporate the full body of inference frameworks \citep[e.g.,][]{moreno-boteBayesianSamplingVisual2011c},
but also makes it biologically more plausible. 
Furthermore, such approaches allow us to make novel predictions, for instance, how uncertainty at different levels of sensory hierarchy \citep{dayanHierarchicalModelBinocular1998}interacts with other cognitive factors, such as reward. 
While extending to more sophisticated Bayesian belief updates is a natural next step, further inspiration can be drawn from recent theoretical developments that cast RL as an inference process \citep{levine2018reinforcement}.

An alternative for integrating Bayesian and RL perspectives is active inference approaches, using a fully Bayesian probabilistic account \citep{parr2022active,whyte2024minimal}. 
Under active inference, for instance, the modulation of binocular rivalry by reward can be explained through the minimization of expected free energy (\nameref{sec:box-1:-predictive}). 
% Expected free energy is decomposed into terms like risk, ambiguity, and novelty. 
The association of one of the rivaling stimuli with reward will minimize the risk term of expected free energy in an additive manner. 
Policies that favor the perception of the rewarded stimulus 
will thus have a lower expected free energy, making them more likely to be selected. 
Consequently, the dominance duration of the rewarded stimulus will be increased (due to active sampling) 
% because the agent is actively sampling information associated with a positive outcome, thus biasing perception towards that stimulus. 
The key objective then is to minimize free energy rather than maximize reward, but the computational architecture of handling uncertainty might be similar.

Both approaches certainly provide promising avenues of investigation \citep[with potential links,][]{friston2009reinforcement,houthooft2016vime,sajid2021active} that help to explain diverse but characteristic pathological behavioral patterns of multistability under a single computational framework. 
This would allow us to bring the decades-old phenomenon of multistability under the same umbrella that we have to study a wide range of cognitive dysfunctions.

\subsection{Neurobiology and computation: two halves of a whole that forgot how to fit}

Behavior, which is the central premise in psychiatry, is supported by a wide range of computations that are realized in the (neuro)physiological machinery of the brain \citep[and body,][]{dayanLevelsAnalysisNeural2006,flavellEmergenceInfluenceInternal2022,safaviBrainComplexSystem2022,tsengDefensiveResponsesBehaviour2023,tian2023evaluation}.
Thus, the underpinnings of psychiatric and neurodevelopmental disorders have been described at both the level of neurobiology and of cognitive computations. For instance, psychopharmacological treatments target the cellular machinery of the brain, while psychotherapeutic interventions target the level of behaviors and computations.
Although mainstream computational psychiatry approaches have focused on cognitive architectures and neural computations, a holistic understanding of psychiatric/neurodevelopmental conditions will require the integration of information processing and neurobiological substrates. We believe the richness of multistability on both sides of this spectrum allows us to build a holistic perspective (\autoref{fig:mspForIntegration}d).

A number of biophysical models of perceptual multistability have focused on its neural dynamics and neurobiological underpinnings, including studies on circuit and synaptic principles such as attractor dynamics, adaptation, and mutual inhibition \citep[e.g.,][]{braunAttractorsNoiseTwin2010,decoBrainMechanismsPerceptual2013,theodoniFluctuationsPerceptualDecisions2014,whyte2025minimal}.
Certain (normative) formulations of perceptual multistability in both schools outlined above also offer synergistic directions, functional and computational interpretations for some traditional biophysical characteristics.
As discussed above and in previous work \citep{leptourgosFunctionalTheoryBistable2020b,moreno-boteBayesianSamplingVisual2011c,dayanHierarchicalModelBinocular1998}, several aspects of the Bayesian account have been connected to underlying neural circuits. 
These include circuit implementation of the hierarchical inference \citep{dayanHierarchicalModelBinocular1998,leptourgos2022hallucinations}, sampling from posterior distributions \citep{moreno-boteBayesianSamplingVisual2011c}, and decision processes 
\citep[][]{caoBinocularRivalryReveals2021a}.
Developing biologically plausible models of RL computations for multistability captured less attention, despite being a promising direction, given the extensive development in the domain of neural RL \citep[][]{dayan2002reward,dayan2008decision,averbeck2022reinforcement} and growing evidence on the involvement of \emph{a large and distributed network} that includes decision-making and action-selection networks \citep{jongIntracranialRecordingsOccipital2016,safaviMultistabilityPerceptualValue2022}.

\subsection{Building synergy between biological and computational psychiatry}

Given that perceptual multistability is observed in a wide variety of species,
it could also be insightful to combine it with animal models of psychiatric disorders. 
This would allow us to connect basic behavioral patterns that are consistent across species to their underlying biological substrate.
In fact, previous invasive studies of multistability in animals, mostly involving non-human primates, have provided valuable insight into the underlying circuit mechanisms \citep[]{leopoldMultistablePhenomenaChanging1999,blakeVisualCompetition2002a,panagiotaropoulosSubjectiveVisualPerception2014b,panagiotaropoulos2024integrative}.
The characteristic hierarchical structure of neural responses observed in the ventral stream of the primate visual system 
was one of the key findings that influenced the inference-based formulation of binocular rivalry \citep{dayanHierarchicalModelBinocular1998,leeHierarchicalBayesianInference2003c}.
Subsequent studies on the role of the prefrontal cortex in perceptual multistability \citep[][]{sterzer2007neural,weilnhammer2013frontoparietal,zaretskayaIntrospectionAttentionAwareness2014,safaviFrontalLobeInvolved2014,odegaardShouldFewNull2017,blockWhatWrongNoReport2019,weilnhammer2021active,michelConsciousPerceptionPrefrontal2022}  contributed to the ignition of the decision-theoretic formulation of perceptual multistability \citep{safaviMultistabilityPerceptualValue2022,martinUsefulMisrepresentationPerception2021,safaviDecisiontheoreticModelMultistability2024}.
Such circuit- and system-level insights connected to cognitive functions can also be helpful to study multistability in animal models of psychiatric or neurodevelopmental conditions, and to understand malfunctioning \emph{circuit mechanisms} --- an approach that has recently gained more momentum \citep{reed2020paranoia,schmackStriatalDopamineMediates2021,suthaharan2024lesions}.

Indeed, recent works on perceptual multistability in preclinical models \citep[e.g.,][]{bogatovaTugofpeaceVisualRivalry2023} have provided results in line with previous work in humans \citep{robertsonSlowerRateBinocular2013a,spiegelSlowerBinocularRivalry2019a} and can open up a vast range of opportunities to understand the underlying impaired biological machinery. In particular, altered switch rates have been reported with mouse models of ASD \citep{palaginaComplexVisualMotion2017,bogatovaTugofpeaceVisualRivalry2023}. 
% compatible with observations in human studies \citep[]{robertsonSlowerRateBinocular2013a,spiegelSlowerBinocularRivalry2019a}.
% Furthermore, behavioral studies in animals and, more importantly, observing similar healthy and pathological behavioral patterns across humans and animals \citep[at least, explicitly in non-human primates and mice][]{palaginaComplexVisualMotion2017,bogatovaTugofpeaceVisualRivalry2023,blakeVisualCompetition2002a,panagiotaropoulosSubjectiveVisualPerception2014b,panagiotaropoulos2024integrative} provide a foundation and a unique opportunity for understanding the phenomena at the circuit level using animal models.
% The recent development of no-report paradigms \citep{tsuchiyaNoReportParadigmsExtracting2015},
% makes multistability an even more appealing paradigm for studies with animals. 
% This highlights the translational potential of multistability for psychiatry. 
% Studying perceptual multistability 
% using animal models of psychiatric conditions opens up a vast range of opportunities to understand the underlying malfunctioning biological machinery from cells to systems.
% For instance, 
Notably, based on existing data and computational models, we can already speculate about the potential circuit mechanisms that lead to altered switch rates. 
% These alterations have been commonly reported both in humans with ASD and its animal models. 
% (as well as in other clinical conditions). 
% As we discussed earlier, the circular inference framework suggests ... . 
For instance, RL and active inference accounts of multistability suggest a potential role for the dopaminergic system in shaping the temporal dynamics of multistability. Previous studies have almost exclusively focused on invasive neural recordings, but there are numerous other avenues of investigation in animals. These include genetics, the study of neuromodulatory systems, pharmacology, and the use of these diverse tools in concert with each other 
\citep{millerGeneticContributionIndividual2010,ngoPsychiatricGeneticStudies2011a,lawEffectStimulusStrength2017,chenGenomicAnalysesVisual2018,carterPsilocybinLinksBinocular2007b,mentchGABAergicInhibitionGates2019}.
Furthermore, the extensive toolbox available for causal interventions (e.g., chemogenetics, optogenetics, etc.), makes animal studies even more appealing.
Thus, given the extensive capacity of combining perceptual multistability with animal research, there is substantial potential to \emph{empirically} connect neural mechanisms to cognitive dysfunctions 
and to bridge human and animal research in psychiatry.

%%% Local Variables:
%%% mode: latex
%%% TeX-master: "../main"
%%% TeX-master: "../main"
%%% End:

%% file: content/conclusionRemark.tex
Here, we argued that the rich and multifaceted nature of multistable perception opens a unique avenue to approach psychiatric and neurodevelopmental disorders, which often involve a wide array of behavioral and neural complexities. 
Interestingly, several aspects of multistability have already been shown to differ across conditions. We discussed these aspects in light of major frameworks in computational psychiatry, including predictive processing and reinforcement learning. 
We also envision several directions where multistability can be leveraged to gain an integrative understanding of psychiatric and neurodevelopmental disorders and synergize with recent advancements in computational psychiatry, such as transdiagnostic approaches and cognitive maps. 
This includes bridging different computational perspectives through novel theoretical frameworks, and connecting two distincts domains in computational psychiatry: neural dynamics and cognitive computations --- also see, \nameref{sec:outstandingQuestions}.
Overall, we believe that perceptual multistability holds significant potential as a promising non-invasive tool for clinical applications.

%%% Local Variables:
%%% mode: latex
%%% TeX-master: "../main"
%%% End:

%% file: content/outstandingQuestions.tex
\begin{highlightbox}
\caption{Outstanding questions}
\label{sec:outstandingQuestions}
Given the essential role of neuromodulatory systems in computations underlying perceptual multistability (inference and decision processes), and their close tie to psychiatric disorders, what is the exact role of different neuromodulatory systems in multistable perception, and multistability-related pathologies?

\qSepSpace

Recent developments in computational psychiatry promoted transdiagnostic analysis as a promising tool to understand common factors across psychiatric and neurodevelopmental disorders, however, such analyses require a large amount of behavioral data. How can we transform classical psychophysical multistability experiments to online settings for massive data collection, and how can such online experiments inform fine-tuned directions for investigating multistability in the context of neuropsychiatric disorders (e.g., in terms of characteristic traits)?

\qSepSpace

Given the potentially rich computational underpinning of perceptual multistability, it is intriguing to ask, how behavioral patterns in multistability are related to behavioral patterns in other cognitive tasks (particularly, the ones central to psychiatry, such as explore-exploit balance, decision-making, working memory, etc.)?

\qSepSpace

Perceptual multistability particularly provides a unique window to understand \emph{internal} cognitive process (e.g., change in perception in the absence of external factors). How can we use multistability as a window to understand the connection to other internal computations that are also central in psychiatry (such as interoception and cognitive maps)?

\qSepSpace

Over the last decades, it has been shown that a large and distributed network (including sensory hierarchy, decision- and value-related networks, etc.) is involved in perceptual multistability. What experiments and computational models do we need to understand the \emph{fine-grained function} of this large and distributed network during multistability and its pathological differences?
\end{highlightbox}

%%% Local Variables:
%%% mode: latex
%%% TeX-master: "../main"
%%% End:

%% file: content/acknowledgment.tex
We would like to thank Christopher Whyte and Peggy Seriès and for their valuable comments on an earlier version of this manuscript.
SS acknowledges the support by Max Planck Society, DFG grant 550411021, and add-on fellowship from the Joachim Herz Foundation. RJ acknowledges the support of the Brain, Society \& Technology Institute (University of Lille), as well as grants from the France2030 PEPR PROPSY program (UNREHAL, ANR-22-EXPR-0006 and CIC-Psy, ANR-22-EXPR-0014). PS acknowledges support from the Swiss National Science Foundation (grant 32003BE\_221944), University of Basel and the Gertrud Thalmann Fonds of the University Psychiatric Clinics (UPK) Basel.
We would like to thank \url{scidraw.io} for providing a free repository of high-quality scientific drawings
(in particular, Akshay Markanday for providing his "Monkey face cartoon" in this repository);
and ``macrovector\_official'' / \href{https://www.freepik.com/}{Freepik} for providing free vector graphics (icons and avatars) that were used in \mfig{fig:mspForIntegration}c.